\documentstyle[multicol,aps,prl]{revtex}

\begin{document}
\draft
\title{{\bf Spontaneous Symmetry Breaking in Directed Percolation with Many Colors:}%
\\
{\bf Differentiation of Species in the Gribov Process}}
\author{{\sc H.K. Janssen}}
\address{{\it Institut f\"{u}r Theoretische Physik III,}\\
Heinrich-Heine-Universit\"{a}t, 40225 D\"{u}sseldorf, Germany}
\date{\today}
\maketitle

\begin{abstract}
A general field theoretic model of directed percolation with many colors
that is equivalent to a population model (Gribov process) with many species
near their extinction thresholds is presented. It is shown that the
multicritical behavior is always described by the well known exponents of
Reggeon field theory. In addition this universal model shows an instability
that leads in general to a total asymmetry between each pair of species of a
cooperative society.
\end{abstract}


\begin{multicols}{2}
\narrowtext

It is well known that the Gribov process \cite{GrSu78,GrTo79} (Reggeon field
theory \cite{Gr67,GM68,Mo78}, stochastic Schl\"{o}gl model \cite{Sch72}) is
a stochastic multiparticle process which describes the essential features of
a vast number of growth phenomena of populations near their extinction
threshold and without exploitation of the environment. The transition
between survival and extinction of the population is a nonequilibrium
continuous phase transition phenomenon and is characterized by universal
scaling laws. The Gribov process belongs to the universality class of local
growth processes with absorbing states \cite{Ja81,Gr82} such as directed
percolation \cite{BrHa57,CaSu80,Ob80}, the contact process \cite
{Ha74,Li85,JeDi94}, and certain cellular automata \cite
{Ki83,Ki85}, and is relevant to a vast range of models in physics,
chemistry, biology, and sociology. I have been studying this transition for
some time by renormalized field theoretic methods \cite{Ja81}. In this
letter I will show that the Gribov process with many interacting species
(directed percolation with many colors) also belongs to this universality
class. In addition this multicolored directed percolation exhibits a new
interesting phenomenon: even if the dynamics of the different colors is
modeled symmetrically, there exists an instability which leads to a
differentiation of the species in their active state. This color symmetry
breaking leads to many different stable forms of a cooperative society.

The multiparticle processes under consideration are characterized by the
following four principles:

\begin{enumerate}
\item  Self-reproduction (``birth'') is error free, but may be different for
each species; particles can be annihilated with rates depending on the
species (``death'').

\item  The particles interact with species-dependent couplings
(``competition'' and ``cooperation'').

\item  Particles diffuse in a $d$-dimensional space (``motion'').

\item  The states where some species are extinct are absorbing.
\end{enumerate}

In the language of chemistry, the system corresponds to an autocatalytic
reaction scheme of the form $X_\alpha \leftrightarrow 2X_\alpha $, $X_\alpha
\rightarrow 0$, $X_\alpha +X_\beta \rightarrow kX_\alpha +lX_\beta $, where
the last reaction subsumes the interactions of species with different colors 
$\alpha $, $\beta ,$ and $k,l=0,1$. A description in terms of particle
densities typically arises from a coarse-graining procedure where a large
number of microscopic degrees of freedom are averaged out. The influence of
these is simply modeled as Gaussian noise-terms in a Langevin equation which
have to respect the absorbing state condition. Stochastic reaction-diffusion
equations for the particle densities $n\left( {\bf r},t\right) =\left(
n_1\left( {\bf r},t\right) ,n_2\left( {\bf r},t\right) ,\ldots \right) $ in
accordance with the four principles given above are constructed as 
\begin{eqnarray}
\partial _tn_a\left( {\bf r},t\right)  &=&\lambda _\alpha \nabla ^2n_\alpha
\left( {\bf r},t\right) +R_\alpha \left( n\left( {\bf r},t\right) \right)
\,n_\alpha \left( {\bf r},t\right)   \nonumber \\
&&+\zeta _\alpha \left( {\bf r},t\right) \ ,  \label{2}
\end{eqnarray}
where the first term on the right hand side models the (diffusive) motion,
and the $R_\alpha $ are the reproduction rates of the particles with color $%
\alpha .$ These deterministic terms are constructed proportional to $%
n_\alpha $ in order to ensure the existence of an absorbing state for each
species. Expanding the rates $R_\alpha $ in powers of $n$ results in 
\begin{equation}
R_\alpha \left( n\left( {\bf r},t\right) \right) =-\lambda _\alpha \left[
\tau _\alpha +\frac 12\sum_\beta g_{\alpha \beta }\,n_\beta \left( {\bf r}%
,t\right) \right]   \label{3}
\end{equation}
up to subleading terms. The Gaussian noises $\zeta _\alpha \left( {\bf r}%
,t\right) $ must also respect the absorbing state condition, whence 
\begin{eqnarray}
\langle \zeta _\alpha \left( {\bf r},t\right) \zeta _\beta \left( {\bf r}%
^{\prime },t^{\prime }\right) \rangle  &=&\lambda _\alpha g_\alpha \delta
_{\alpha ,\beta }\,n_\alpha \left( {\bf r},t\right)   \nonumber \\
&&\times \delta \left( {\bf r}-{\bf r}^{\prime }\right) \delta \left(
t-t^{\prime }\right)   \label{4}
\end{eqnarray}
also up to subleading contributions. In this letter I only consider
color-independent diffusion constants $\lambda _\alpha =\lambda $. Then, for
stability, it must be demanded $\sum_{\alpha ,\beta }n_\alpha g_{\alpha
\beta }n_\beta \geq 0$ for all $n_\alpha \geq 0.$ The ``temperature''
variables $\tau _\alpha $ measure the difference of the rates of death and
birth of the species $\alpha $. Thus the temperatures may be positive ore
negative. I am interested in the case where all $\tau _\alpha \approx 0$
(up to fluctuation corrections) defining the multicritical region. Under
these conditions all the species live on the border of extinction.

The equations of motion (\ref{2}) show (absorbing) homogeneous steady state
solutions with any combination of extinct species. In the infinite volume
limit one finds transitions to steady states with $n_\alpha \geq 0$ which
are given in mean field theory as the solutions of $\sum_\beta \,g_{\alpha
\beta }n_\beta =-2\tau _\alpha $ if $n_\alpha >0$. This set of equations in
general offers multiple different solutions if some or all of the $\tau
_\alpha $ are negative. Each of these solutions defines a different
homogeneous phase of the active system. In some of these phases only one of
the species is alive, in the other phases several species act cooperatively.
In the space spanned by the temperature variables $\tau _{\alpha {}}$, the
phases are separated by surfaces of first or second order transitions where
a species becomes extinct. All the critical surfaces meet together in a
multicritical point where all temperatures are zero. Take e.g.\ the simplest
case of a totally symmetric three-species system with $g_{\alpha \beta }=g$
and $\tau _\alpha <0$ for $\alpha ,\beta =1,2,3$. The ternary phase diagram
is shown in Fig.1. In this case there only exist phases where one species is
nonzero and the two others are suppressed. Here stable cooperative phases
with more than one species are not possible.

In general, this mean field picture continuously depends on the coupling
constants $g_{\alpha \beta }$. In the following I will show that in spatial
dimensions lower than $4$, the stochastic equations (\ref{2}-\ref{4} )
produce universal macroscopic steady state equations which correspond to
only a discrete set of choices for the coupling matrix ${\bf g}$ that I call
strategies. The stable ones are distinguished by a total asymmetry between
the members of each pair of species, in a mean field picture: $g_{\alpha
\beta }$ or $g_{\beta \alpha }$ is zero. This implies that only one of these
partners directly influences the evolution of the other. These stable
strategies with differentiated species permit cooperative phases with more
than one species nonzero and continuous phase transitions. All the
continuous phase transitions of each strategy are characterized by the
directed percolation critical exponents of the one-species Gribov process.

In the following I will be interested in the behavior of correlation- and
response functions in the vicinity of the multicritical point where all $%
\tau _\alpha $ are small. The critical properties are extracted using
renormalized field theory \cite{Am84,ZiJu93,BJW76,DDP78,Ja79,Ja92} in
conjunction with an $\varepsilon $-expansion about the upper critical
dimension $d_c=4$. It is then convenient to recast the Langevin equation as
a dynamical functional \cite{DeDo76,Ja76} 
\begin{eqnarray}
{\cal J}\left[ \widetilde{s},s\right]  &=&\int dt\,d^dr\,\sum_\alpha \left\{ 
\widetilde{s}_\alpha \left[ 
\rule[0.0in]{0.0in}{0.3in}%
\partial _t+\lambda _\alpha \left( \tau _\alpha -\nabla ^2\right) \right.
\right.   \nonumber \\
&&\left. \left. +\frac 12\sum_\beta g_{\alpha \beta }s_\beta -\frac 12%
g_\alpha \widetilde{s}_\alpha \right] s_\alpha \right\} \ ,  \label{6}
\end{eqnarray}
where $s_\alpha \left( {\bf r},t\right) \sim n_\alpha \left( {\bf r}%
,t\right) $ are rescaled density variables which ensure that $g_{\alpha
\alpha }=g_\alpha $. The $\widetilde{s}_\alpha \left( {\bf r},t\right) $
denote the response fields (the MSR auxiliary variables \cite{MSR72}).
Correlation- and response functions can now be expressed as functional
averages of monomials of $s$ and $\tilde{s}$ with weight $\exp \left\{ -%
{\cal J}\right\} .$ The responses are defined with respect to additional
local particle sources $\widetilde{h}_\alpha \left( {\bf r},t\right) \geq 0$
in the equation of motion (\ref{2}). By these sources a further term $%
-\sum_\alpha \widetilde{h}_\alpha \widetilde{s}_\alpha $ is added to the
integrand of the dynamic functional ${\cal J}$ (\ref{6}). Coupling constants
invariant under the rescaling $s_\alpha \rightarrow c_\alpha s_\alpha ,\ 
\widetilde{s}_\alpha \rightarrow c_\alpha ^{\,-1}\widetilde{s}_\alpha $ are
given by $f_{\alpha \beta }=g_{\alpha \beta }g_\beta $. The scaling by the
usual convenient length $\mu ^{-1}$ and time scale $\left( \lambda \mu
^2\right) ^{-1}$, respectively, shows $\widetilde{s}_\alpha \sim s_\alpha
\sim \mu ^{d/2},$ $f_{\alpha \beta }\sim \mu ^\varepsilon $ where $%
\varepsilon =4-d$ which signals $d_c=4$ to be the upper critical dimension.
A further inspection of (\ref{6}) and a glance at the perturbation expansion
shows that the functional ${\cal J}$ includes all relevant couplings and is
thus renormalizable.

To determine the renormalizations, one has to calculate the primitively
divergent vertex functions $\Gamma _{\alpha ,\alpha }$\thinspace , $\Gamma
_{\alpha \alpha ,\alpha }=\Gamma _{\alpha ,\alpha \alpha }$\thinspace , $%
\Gamma _{\alpha ,\alpha \beta }$ with nonnegative $\mu $-dimensions. Here
the first and the last part of indices denotes the amputated $\widetilde{s}$%
- and $s$-legs, respectively. Dimensional regularization and minimal
renormalization is used. The $\varepsilon $-poles are absorbed by
multiplicative renormalizations of the fields and couplings which I choose
in the form 
\begin{eqnarray}
s_\alpha  &\rightarrow &%
\mathaccent"7017{s}%
_\alpha =\sqrt{Z_s^{(\alpha )}}\,s_\alpha \ ,\quad \widetilde{s}_\alpha
\rightarrow 
\mathaccent"7017{\widetilde{s}}%
_\alpha =\sqrt{Z_s^{(\alpha )}}\,\widetilde{s}_\alpha \ ,  \nonumber \\%
[0.25cm]
\lambda _\alpha  &\rightarrow &%
\mathaccent"7017{\lambda}%
_\alpha =\frac{Z_\lambda ^{(\alpha )}}{Z_s^{(\alpha )}}\,\lambda _\alpha \
,\quad \tau _\alpha \rightarrow 
\mathaccent"7017{\tau}%
_\alpha =\frac{Z_\tau ^{(\alpha )}}{Z_\lambda ^{(\alpha )}}\,\lambda _\alpha
\ ,  \nonumber \\[0.25cm]
f_{\alpha \beta } &\rightarrow &%
\mathaccent"7017{f}%
_{\alpha \beta }=G_\varepsilon ^{\,-1}\mu ^\varepsilon \,\frac{Z_u^{(\alpha
\beta )}}{Z_s^{(\beta )}Z_\lambda ^{(\alpha )}Z_\lambda ^{(\beta )}}%
\,u_{\alpha \beta }\ .  \label{7}
\end{eqnarray}
Here $G_\varepsilon =\Gamma \left( 1+\varepsilon /2\right) /\left( 4\pi
\right) ^{d/2}$ is a convenient constant. From the topology of the diagrams
of the perturbation expansion it is easily seen that the vertex functions do
not depend on coupling constants with indices other than the ones of the
vertex functions itself. Therefore the renormalization factors $Z_i^{(\alpha
)}$ with $i=\left( s,\tau ,\lambda \right) $ and $Z_u^{(\alpha
)}:=Z_u^{(\alpha \alpha )}$ that are determined from $\Gamma _{\alpha
,\alpha }$ and $\Gamma _{\alpha \alpha ,\alpha }$ are already known from the
simple Gribov process. Thus the renormalization group flows of the fields
and parameters with exception of the $g_{\alpha \beta }$ with $\alpha \neq
\beta $ do not depend on the coupling between different colors. It follows
at once that all the critical exponents are given by the usual directed
percolation exponents and the fixed point values of the ``diagonal''
renormalized couplings $u_{\alpha \alpha }=u_\alpha $ are all independent
from the color $\alpha .$

The flow of the couplings between different colors follows from the $%
Z_u^{(\alpha \beta )}$ with $\alpha \neq \beta $. The symmetric fixed point
might be attractive. To disprove that I have calculated these
renormalizations up to two loop order for the special case $\lambda _\alpha
=\lambda $, $u_\alpha :=u_{\alpha \alpha }=u$ with the result

\begin{eqnarray*}
Z_u^{\left( \alpha \beta \right) } &=&1+\frac 1{4\varepsilon }\left(
6u+u_{\alpha \beta }+u_{\beta \alpha }\right) +\frac 1{16\varepsilon }\left[
\left( \frac{36}\varepsilon -\frac{19}2\right) u^2\right.  \\[0.1cm]
&&\left. +\left( \frac 8\varepsilon -\frac 54\right) uu_{\beta \alpha
}+\left( \frac 1\varepsilon -\frac 32\ln \frac 43\right) \left( u_{\alpha
\beta }^2+u_{\beta \alpha }^2\right) \right.  \\
&&\left. +\left( \frac 2\varepsilon -1\right) u_{\alpha \beta }u_{\beta
\alpha }+\left( \frac 8\varepsilon -\frac 94+3\ln \frac 43\right) u_{\alpha
\beta }u\right] .
\end{eqnarray*}
Collecting together these and the known results for the other
renormalization factors, one easily gets the flow equation for the
renormalized coupling constants 
\begin{equation}
\mu \frac \partial {\partial \mu }u_{\alpha \beta }=\beta _{\alpha \beta }
\label{9}
\end{equation}
with the Wilson-functions 
\begin{eqnarray}
\beta _{\alpha \beta } &=&\left[ -\varepsilon +u+\frac 14\left( u_{\alpha
\beta }+u_{\beta \alpha }\right) -\frac 18u_{\alpha \beta }u_{\beta \alpha
}\right.   \nonumber \\[0.13cm]
&&\left. -\left( \frac{97}{128}+\frac{53}{64}\ln \frac 43\right) u^2-\left( 
\frac 9{32}-\frac 38\ln \frac 43\right) uu_{\alpha \beta }\right.   \nonumber
\\
&&\left. -\frac 5{32}uu_{\beta \alpha }-\frac 3{16}\ln \frac 43\left(
u_{\alpha \beta }^2+u_{\beta \alpha }^2\right) +\ldots \right] u_{\alpha
\beta }\ .  \label{10}
\end{eqnarray}

For $\alpha =\beta $, the equation $\beta _{\alpha \alpha }=:\beta \left(
u_\alpha \right) =0$ yields the nontrivial stable fixed point value $u_{*}=%
\frac 23\varepsilon \left\{ 1+\left( \frac{169}{288}+\frac{53}{144}\ln \frac %
43\right) \varepsilon +\ldots \right\} $. For each pair of species with
colors $\alpha \neq \beta $, fixed point solutions of (\ref{9}) with at
least one zero coupling are either given by $u_{\alpha \beta *}=u_{\beta
\alpha *}=0$, or by the pair of solutions $u_{\alpha \beta *}=\left(
1+\varepsilon _{\alpha \beta }\right) w_{*}$ with $\varepsilon _{\alpha
\beta }=-\varepsilon _{\beta \alpha }=\pm 1$ and $w_{*}=u_{*}+O\left(
\varepsilon ^3\right) $. To analyze the case with $u_{\alpha \beta *}$ and $%
u_{\beta \alpha *}$ both nonzero, one sets $u_{\alpha \beta
*}=u_{*}+v_{\alpha \beta }$. To first order in $\varepsilon $ one finds a
fixed point line $v_{\alpha \beta }+v_{\beta \alpha }=0.$ But using the
higher order terms of (\ref{10}), one gets $v_{\alpha \beta }=0+O\left(
\varepsilon ^2\right) $. Then it can be easily seen that $v_{\alpha \beta }$
must be zero order for order in the $\varepsilon $-expansion. This proves
that in this case the fixed point is indeed symmetric $u_{\alpha \beta
*}=u_{\beta \alpha *}=u_{*}$. Thus I have found from this analysis that for
each pair of species independent of the other ones three types of fixed
points for the couplings exist: both couplings vanish (decoupled), or only
one of the couplings is nonzero (asymmetric), or both couplings are nonzero
and equal to the diagonal one (symmetric).

The stability of the various fixed points is analyzed by the calculation of
the eigenvalues of the matrix of derivatives of the Wilson-functions $\beta
_{\alpha \beta },\ \beta _{\beta \alpha }.$ This analysis yields as stable
fixed points only the asymmetric ones. The decoupled fixed points are doubly
instable, the two instability exponents are equal to $\omega _d=\frac 12%
u_{*}\left( 1-\frac 98u_{*}\right) +\ldots =\frac \varepsilon 3\left(
1-\varepsilon \,0.0573+\ldots \right) $. The symmetric fixed points have one
instability exponent $\omega _s=\left( 1-3\ln \frac 43\right)
u_{*}^{\,2}+\ldots =\varepsilon ^2\,0.0609+\ldots $. I especially note the
very slow crossover from the symmetric to the asymmetric fixed point.

To get a simple qualitative impression of the behavior of the dynamic system
at its asymmetric fixed points, I consider, in the active region with all $%
-\tau _\alpha =:g\,r_\alpha \geq 0$, a renormalized mean-field theory of the
equations (\ref{2},\ref{3}) for spatially homogeneous densities $\bar{n}_a$%
and set $g_{\alpha \alpha }=g$, $g_{\alpha \beta }=g\left( 1+\varepsilon
_{\alpha \beta }\right) $ and $\sum_\alpha \bar{n}_a=\bar{n}$: 
\begin{equation}
\partial _t\bar{n}_a=\lambda g\left( r_\alpha -%
{\textstyle {1 \over 2}}%
\bar{n}-%
{\textstyle {1 \over 2}}%
\sum_\beta \varepsilon _{\alpha \beta }\,\bar{n}_\beta \right) \,\bar{n}%
_\alpha \ .  \label{12}
\end{equation}
The decision which of all the $\varepsilon _{\alpha \beta }=-\varepsilon
_{\beta \alpha }$ are $\pm 1$ defines a stable strategy. The strategies may
be pictured by a strategy-diagram. Each corner of such a diagram represents
a color $\alpha $. Consider now each pair $\left( \alpha ,\beta \right) $ of
corners, they are connected by a line with an arrow pointing at $\alpha $ if 
$\varepsilon _{\alpha \beta }=1$ or vice versa if $\varepsilon _{\alpha
\beta }=-1$. The arrows therefore represent the ``pressure'' on the
reproduction rate resulting from one species to another. The diagrams for
the stable strategies are therefore fully connected by such directed lines.
For a system with three species e.g.\ two stable strategies are possible, a
cyclic and a hierarchic one (upper part of Fig.2). In (\ref{12}) the
influence on the reproduction rates of the society as a whole is described
by the common ``pressure'' $p=g\bar{n}/2$. If now $\varepsilon _{\alpha
\beta }=1$ e.g., the part of $p$ resulting from the species $\beta $ on the
reproduction of $\alpha $ is doubled whereas the part of the common pressure
resulting from the species $\alpha $ on the reproduction of $\beta $ is
fully compensated. As a result there exist many cooperative phases with more
then one species as stationary solutions of the mean field equations of
motion. As an example the phase diagrams for the stable strategies of a
system of three species is shown in Fig.2 (lower part).

Now the unstable strategies can be pictured as diagrams with disconnected
corners for a pair $\left( \alpha ,\beta \right) $ in the uncoupled case $%
g_{\alpha \beta }=0$ (then there is no pressure from $\alpha $ on $\beta $
and vice versa), and connected corners but without direction for a pair $%
\left( \alpha ,\beta \right) $ in the symmetric case $g_{\alpha \beta
}=g_{\beta \alpha }$ (then the species $\alpha ,\beta $ do influence their
reproduction rates only by means of the common pressure $p$). It is easily
seen that in the fully symmetric case (for three species see Fig.1) more
then one species can coexist only if their temperatures $\tau _\alpha
=-g\,r_\alpha $ are all equal. If for one color $\alpha $ the parameter $%
\,r_\alpha $ is slightly increased, the species $\alpha $ suppresses all the
others, if $\,r_\alpha $ is decreased, it becomes extinct by the common
pressure of all the others.

Summarily I have shown that also in stochastic multispecies models of
populations that develops near the extinction threshold of all the species
and therefore have many absorbing states, the critical properties at the
multicritical point and at all continuous transitions are governed by the
well known Gribov process (Reggeon field theory) exponents (for a previous
simulational result on a two-species system which seems to agree with my
findings see \cite{CDDM91}). In other regions of the phase diagram of course
more complicated critical behavior may arise such as multicritical points
with different scaling exponents (see the recent work of Bassler and Brown 
\cite{BaBr96}). The models considered here have many absorbing states (each
combination of species may be extincted irrespectively of the other ones)
and therefore are different from models which were considered by Grinstein
et al.\ \cite{GLB89} where it was shown that multispecies systems with one
absorbing state belong to the Gribov universality class. In addition I have
shown that the universal properties of interspecies correlations and the
phase diagram are determined by totally asymmetric fixed point values of the
renormalized interspecies coupling constants. This eventually leads to a
system working cooperatively. It is interesting that the asymmetry between
the pairs of species seems to be the condition for this cooperation near
extinction. The model considered here is a simple but universal model of
such a cooperative society and should therefore have many applications in
all fields of natural science up to sociology.

\acknowledgments I thank S.\ Theiss for a critical reading of the
manuscript. This work has been supported in part by the SFB 237 of the
Deutsche Forschungsgemeinschaft.

%
\begin{figure}
\caption{Symmetric strategy of a three-species-system and its ternary phase diagram
 in the plane $ \tau_{1} + \tau_{2} + \tau_{3} = -r < 0 $. The numbers $ \alpha $
 at the corners denote the points where the corresponding $ \tau_{\alpha} = -r $
 and the other temperatures are zero. Dashed lines indicate discontinuous transitions
 between the one-species phases.}
\end{figure}

\begin{figure}
\caption{Cyclic and hierarchic strategy of a three-species system and their ternary
 phase diagrams. The numbers inside the phase diagrams denote the colors of
 the living species. Solid lines indicate continuous transitions.}
\end{figure}
%

\end{multicols}

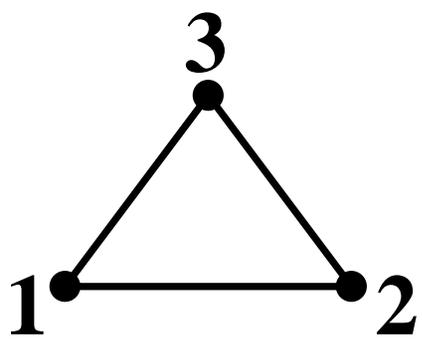
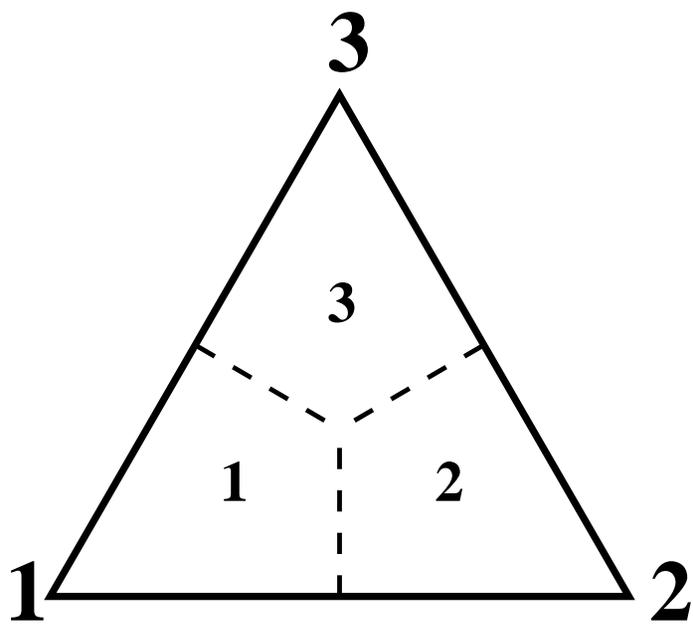
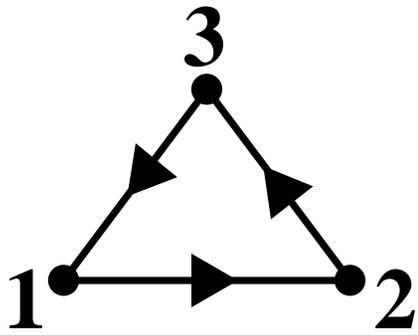
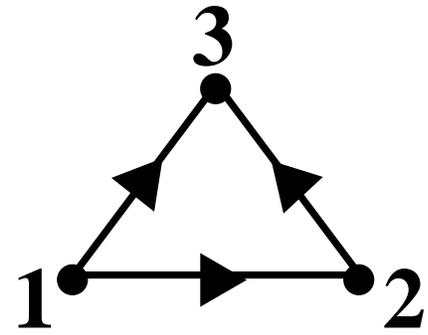
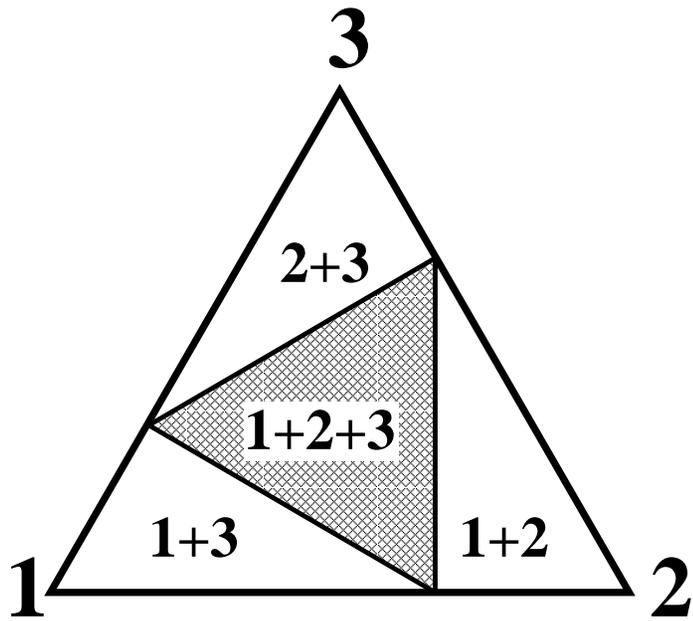
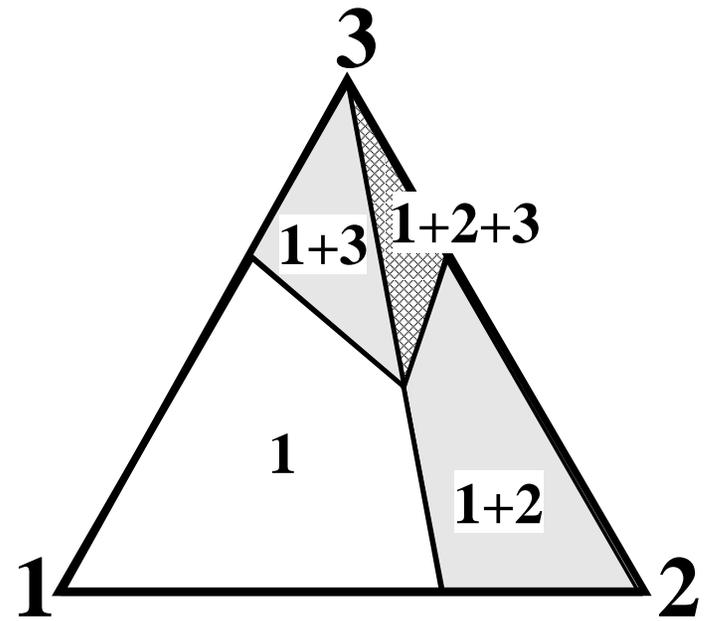
\end{document}